\documentclass[preprint,5p,times,twocolumn]{elsarticle}

\usepackage{amsmath,amssymb}
\usepackage{graphicx}
\usepackage{hyperref}
\usepackage{slashed}
\usepackage[utf8]{inputenc}
\usepackage{textalpha}
\journal{Physics Letters B}

\begin{document}

\begin{frontmatter}

\title{Neutrino Portal to Extra Dimensions:\\ Unified Origin of Dark Radiation, Dark Matter, and Neutrino Decay}

\author{Arnab Chaudhuri\corref{cor1}}
\address{Division of Science, National Astronomical Observatory of Japan, Mitaka, Tokyo 181-8588, Japan}
\cortext[cor1]{Corresponding author. Email: \texttt{arnab.chaudhuri@nao.ac.jp}}

\begin{abstract}
We propose a unified framework based on sterile neutrinos propagating in large extra dimensions (LED) and coupled to a pseudo-Nambu-Goldstone boson (Majoron) from spontaneous lepton number violation. In this setup, the lightest Kaluza–Klein (KK) sterile neutrino serves as warm dark matter, higher KK modes decay invisibly producing a relativistic Majoron contributing to $\Delta N_{\rm eff}$, and active neutrinos undergo invisible decay. This model addresses dark radiation, dark matter, and neutrino anomalies within a minimal, testable framework. We present the Boltzmann evolution of relevant species, identify the viable parameter space, and highlight implications for CMB-S4, Lyman-$\alpha$, and neutrino observatories.
\end{abstract}


\end{frontmatter}


\section{Introduction}

The $\Lambda$CDM paradigm has been remarkably successful in explaining a wide range of observations, including the cosmic microwave background (CMB) anisotropies, large-scale structure formation, and Big Bang nucleosynthesis (BBN) \cite{Planck:2018vyg, Aghanim:2018eyx, BOSS:2016wmc}. However, over the past decade, various small but persistent tensions have emerged, motivating theoretical extensions of the Standard Model (SM) and its cosmological embedding. Among these are mild excesses in the inferred value of the effective number of relativistic degrees of freedom ($\Delta N_{\rm eff}$) from CMB and BBN \cite{Fields:2019pfx, Cyburt:2015mya}, the small-scale crisis in structure formation \cite{Bullock:2017xww, Weinberg:2013aya}, and hints of non-standard neutrino interactions or invisible decays \cite{Kasieczka:2022naq, Chacko:2019nej, IceCube:2020fpi}. These observations suggest the presence of new feebly interacting particles that were never in thermal equilibrium, but subtly shaped cosmological evolution.

A compelling class of candidates arises from the physics of sterile neutrinos and pseudo-Nambu-Goldstone bosons (pNGBs), particularly the Majoron \cite{Chikashige:1980ui, Gelmini:1980re}. Majorons emerge from the spontaneous breaking of a global lepton number symmetry and provide natural channels for neutrino mass generation and decay. When sterile neutrinos couple to the Majoron, their decays can inject relativistic particles into the plasma, modifying the radiation content of the Universe. Simultaneously, if the lightest sterile state is stable, it may act as a warm dark matter (WDM) candidate \cite{Boyarsky:2018tvu, Abazajian:2017tcc}, contributing to resolving the small-scale structure problems without violating Lyman-$\alpha$ bounds \cite{Irsic:2017ixq}.

Large extra dimensions (LED) offer a compelling framework to unify these ingredients. In LED models, SM singlets such as sterile neutrinos can propagate in a higher-dimensional bulk \cite{Arkani-Hamed:1998wuz, Dienes:1998sb}, giving rise to a tower of Kaluza–Klein (KK) excitations in the effective four-dimensional theory. The smallness of active neutrino masses can be attributed to volume suppression, while the KK tower structure naturally introduces a hierarchical spectrum of sterile states with decreasing lifetimes and increasing masses. The decay of heavy KK modes into lighter states and Majorons, as well as the invisible decay of active neutrinos, leads to a cosmological history enriched by both dark matter and dark radiation production \cite{Mohapatra:1999af, Dvali:1999cn, Barbieri:2000mg}.

In this work, we propose a minimal and predictive extra-dimensional model where the sterile neutrino KK tower is coupled to a light Majoron, either localized on the brane or weakly spread in the bulk. The lightest KK sterile neutrino $\nu^{(1)}$ is produced via freeze-in from Higgs or scalar decays and survives as a viable WDM candidate \cite{Hall:2009bx, McDonald:2001vt}. The heavier KK states decay to active neutrinos and relativistic Majorons, contributing to $\Delta N_{\rm eff}$ in a calculable manner \cite{Frigerio:2011in, DEramo:2020gpr}. The invisible decay of active neutrinos, driven by small Yukawa couplings, becomes observable in high-energy cosmic neutrino observatories \cite{Beacom:2002vi, Bustamante:2020mep} and upcoming long-baseline experiments such as DUNE \cite{DUNE:2020ypp}.

We present a comprehensive study of the Boltzmann evolution of all relevant species in this model, including sterile neutrino production, Majoron radiation injection, and neutrino decay. We delineate the viable parameter space consistent with bounds from the CMB \cite{Planck:2018vyg}, BBN \cite{Fields:2019pfx}, Lyman-$\alpha$ forest \cite{Irsic:2017ixq}, supernova cooling \cite{Raffelt:1990yz, Kachelriess:2000qc}, and laboratory neutrino experiments \cite{Kasieczka:2022naq}. The model predicts correlated signals in $\Delta N_{\rm eff}$ and invisible neutrino decay rates, accessible in next-generation CMB probes such as CMB-S4 \cite{CMB-S4:2016ple}, Simons Observatory \cite{SimonsObservatory:2018koc}, and neutrino detectors like IceCube  \cite{IceCube:2020fpi}. This unified framework provides an economical yet rich structure capable of addressing several long-standing anomalies in modern cosmology and neutrino physics.

\section{Model Framework}

We consider a minimal and predictive extension of the Standard Model (SM) wherein a single SM-singlet fermion $\nu_R(x, y)$ propagates in a $D = 4 + \delta$ dimensional spacetime, with the $\delta$ extra spatial dimensions compactified on a torus $T^\delta$ of common radius $R$. All SM fields are confined to a four-dimensional 3-brane, while the right-handed neutrino $\nu_R$ is allowed to propagate freely in the bulk. This setup follows the large extra dimension (LED) paradigm, which offers a natural solution to the neutrino mass hierarchy problem~\cite{Arkani-Hamed:1998wuz, Dienes:1998sb, Mohapatra:1999af}.

Upon compactification, the bulk neutrino field can be Fourier-expanded into a tower of four-dimensional Kaluza--Klein (KK) modes:
\begin{equation}
\nu_R(x, y) = \sum_{\vec{n}} \frac{1}{\sqrt{V_\delta}} \, \nu_R^{(\vec{n})}(x) \, e^{i \vec{n} \cdot \vec{y}/R},
\end{equation}
where $\vec{n} \in \mathbb{Z}^\delta$ labels the KK excitation levels, and $V_\delta = (2\pi R)^\delta$ denotes the compactification volume. Each KK mode acquires a mass through the usual relation
\begin{equation}
m_n = \sqrt{m_0^2 + \frac{|\vec{n}|^2}{R^2}},
\end{equation}
where $m_0$ is a brane-localized Dirac mass induced via the Yukawa interaction $y_\nu \bar{L} H \nu_R(x, y=0) + \text{h.c.}$ The smallness of active neutrino masses arises naturally through volume suppression in this framework, yielding $m_\nu \sim y_\nu v / \sqrt{V_\delta}$~\cite{Dvali:1999cn}.

To further enrich the phenomenology, we introduce a pseudo-Nambu--Goldstone boson (pNGB) — the \emph{Majoron} $J$ — arising from the spontaneous breaking of a global lepton number symmetry $U(1)_L$ at a scale $f$~\cite{Chikashige:1980ui, Gelmini:1980re}. Depending on the UV completion, the Majoron may be brane-localized or propagate into a subset of the bulk. Its interaction with neutrino mass eigenstates in the effective theory is described by
\begin{equation}
\mathcal{L}_{\rm int} \supset \frac{g_0}{\sqrt{V_\delta}} \sum_{\vec{n}, i} \nu_L^i \nu_R^{(\vec{n})} J + \text{h.c.},
\end{equation}
where $g_0$ is a dimensionless coupling of order unity in the higher-dimensional theory. The effective four-dimensional couplings $g^{(n)}$ are suppressed by the compactification volume, and their precise values also depend on the localization profile of $J$. Typically, higher KK modes possess smaller effective couplings due to either volume dilution or wavefunction orthogonality.

This framework leads to a hierarchical tower of sterile neutrino states, with increasing mass and decreasing lifetime for higher KK levels. The lightest KK mode $\nu_R^{(1)}$ is assumed to be cosmologically stable and can serve as a viable warm dark matter (WDM) candidate. Its production occurs via the freeze-in mechanism through suppressed Higgs decays $H \to \nu_L \nu_R^{(1)}$, and it never reaches thermal equilibrium with the SM plasma~\cite{Hall:2009bx, Kusenko:2010ik, Merle:2013wta}.

Heavier KK modes $\nu_R^{(n)}$ with $n > 1$ are unstable and decay predominantly into active neutrinos and relativistic Majorons, $\nu_R^{(n)} \to \nu_L + J$, contributing to the dark radiation component of the Universe and modifying the effective number of relativistic species $\Delta N_{\rm eff}$~\cite{Frigerio:2011in, Escudero:2019gvw}. The decays typically occur after the KK modes freeze out or are produced via higher-order freeze-in processes, and the timing of the decay critically determines their contribution to cosmological observables such as the CMB and BBN.

In addition, the coupling of the Majoron to the active neutrino sector enables the invisible decay channel $\nu_i \to \nu_j + J$, where $i > j$ denotes the mass hierarchy. These decays are constrained by neutrino lifetime bounds from solar and atmospheric experiments, as well as from high-energy neutrino fluxes observed at IceCube~\cite{Beacom:2002vi, Bustamante:2020mep, Barenboim:2020vrr}. The decay width scales as $\Gamma \propto |g_{ij}|^2 m_{\nu_i}$, making even small couplings relevant for cosmological timescales.

Taken together, the combination of sterile KK neutrinos, the Majoron portal, and the LED framework yields a rich cosmological history. This setup provides a unified origin for:
\begin{itemize}
    \item Warm dark matter from freeze-in production of $\nu_R^{(1)}$,
    \item Dark radiation from late decays of higher KK states into Majorons,
    \item Non-standard neutrino decay signatures in terrestrial and astrophysical data.
\end{itemize}

\begin{figure}[t]
    \centering
    \includegraphics[width=0.48\textwidth]{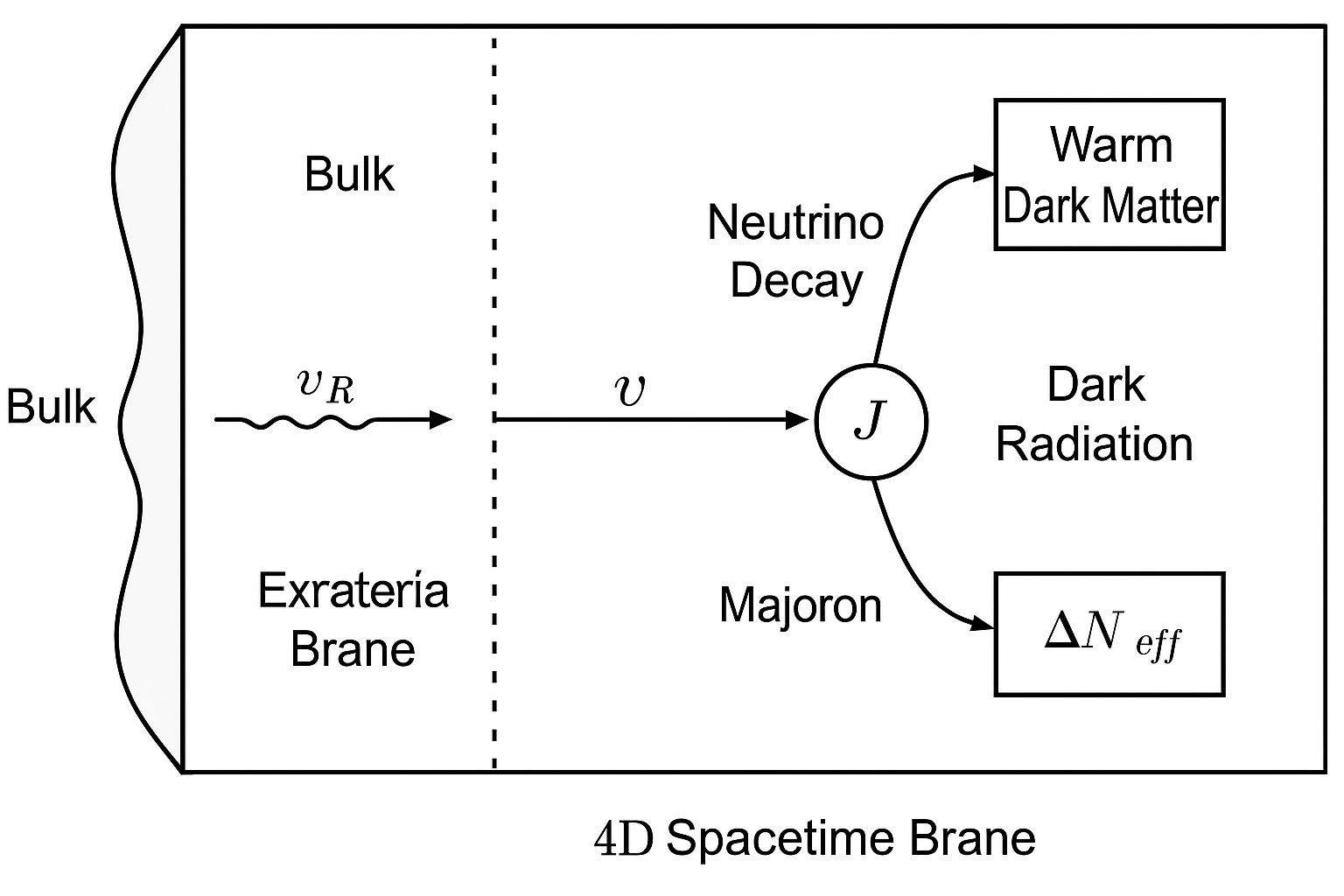}
    \caption{Schematic representation of the setup: a bulk right-handed neutrino $\nu_R$ propagates through $\delta$ extra dimensions and couples to SM neutrinos $\nu$ on the 4D spacetime brane. The interaction is mediated by a Majoron $J$, leading to neutrino decay, dark radiation (contributing to $\Delta N_{\rm eff}$), and warm dark matter production.}
    \label{fig:schematic}
\end{figure}

This is schematically illustrated in Fig.~\ref{fig:schematic}, where the interplay between extra-dimensional propagation, Majoron-mediated interactions, and SM brane-localization drives the correlated phenomenology. In what follows, we analyze the cosmological consequences of this model, including the calculation of the freeze-in abundance, Boltzmann evolution of radiation injection, and the constraints imposed by $\Delta N_{\rm eff}$ and neutrino lifetime bounds.

\section{Dark Matter and Radiation Production}

In the present framework, the lightest Kaluza–Klein (KK) mode of the sterile neutrino, $\nu_R^{(1)}$, plays the role of a warm dark matter (WDM) candidate. Unlike thermal relics, its abundance is determined via the freeze-in mechanism, wherein $\nu_R^{(1)}$ is slowly populated through the feeble decay of Standard Model (SM) particles without ever reaching thermal equilibrium with the plasma. The dominant production channel arises from the decay of the Higgs boson:
\begin{equation}
H \to \nu_L \, \nu_R^{(1)}.
\end{equation}

Assuming a suppressed Yukawa interaction $y \ll 1$, the partial decay width for this channel is given by
\begin{equation}
\Gamma(H \to \nu \nu^{(1)}) = \frac{y^2 m_H}{8\pi} \left(1 - \frac{4 m_{\nu^{(1)}}^2}{m_H^2} \right)^{3/2},
\end{equation}
where $m_H$ is the Higgs boson mass, and $m_{\nu^{(1)}}$ is the mass of the lightest KK mode.

The freeze-in production yield is governed by the Boltzmann equation, accounting for the decays of thermally distributed Higgs bosons:
\begin{equation}
\frac{dY_{\nu^{(1)}}}{dT} = \frac{135 \sqrt{10}}{4 \pi^7 g_*^{3/2}} \frac{M_{\rm Pl} \Gamma_H}{T^6} K_1\left(\frac{m_H}{T}\right),
\end{equation}
where $M_{\rm Pl}$ is the Planck mass, $g_*$ is the effective number of relativistic degrees of freedom at temperature $T$, and $K_1$ is the modified Bessel function of the second kind. 

Integrating over temperature from reheating down to the freeze-in decoupling temperature yields an approximate analytic expression for the final yield:
\begin{equation}
Y_{\nu^{(1)}}^{\rm FI} \approx \frac{135 \sqrt{10} \Gamma_H M_{\rm Pl}}{4 \pi^7 g_*^{3/2} m_H^5}.
\end{equation}

The relic abundance today is related to the freeze-in yield by
\begin{equation}
\Omega h^2 = 2.75 \times 10^8 \left( \frac{m_{\nu^{(1)}}}{\rm GeV} \right) Y_{\nu^{(1)}},
\end{equation}
which imposes a strong constraint on the combination $y^2 / m_{\nu^{(1)}}$ in order to achieve the observed dark matter abundance $\Omega_{\rm DM} h^2 \approx 0.12$.

The same framework predicts a tower of heavier sterile KK modes, $\nu_R^{(n)}$ with $n > 1$, which decay into lighter states via the Majoron channel:
\begin{equation}
\nu_R^{(n)} \to \nu + J.
\end{equation}
These decays are governed by the effective coupling $g^{(n)}$ and proceed at a rate
\begin{equation}
\Gamma_{\nu^{(n)} \to \nu + J} \simeq \frac{|g^{(n)}|^2}{16\pi} m_{\nu^{(n)}},
\end{equation}
producing relativistic Majorons that contribute to the dark radiation component of the Universe. The resulting injection of non-thermal radiation has a measurable impact on the effective number of neutrino species, $\Delta N_{\rm eff}$, and is therefore tightly constrained by cosmic microwave background (CMB) and big bang nucleosynthesis (BBN) data. 

\begin{figure*}[t]
    \centering
    \includegraphics[width=0.48\textwidth]{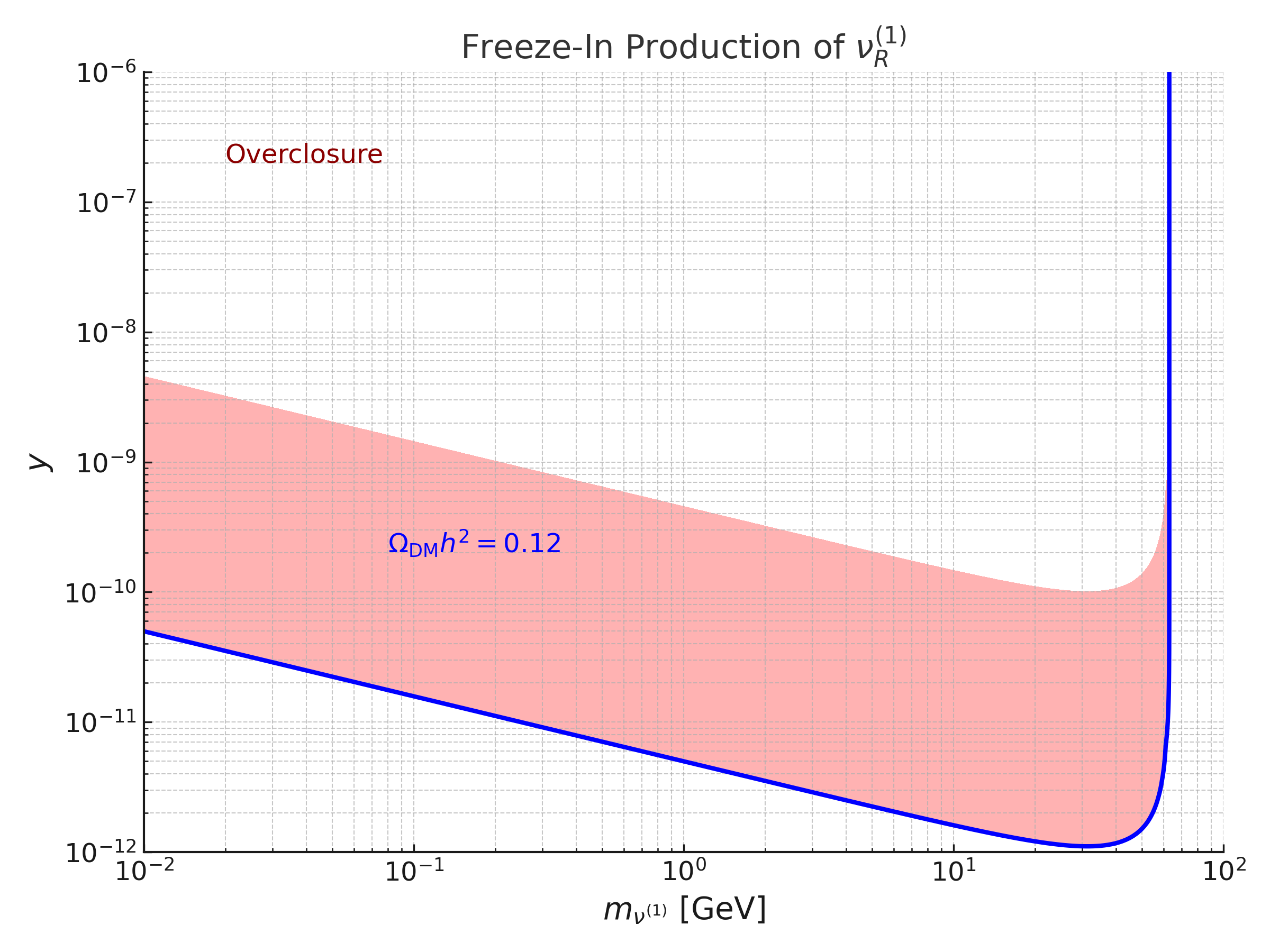}
    \includegraphics[width=0.48\textwidth]{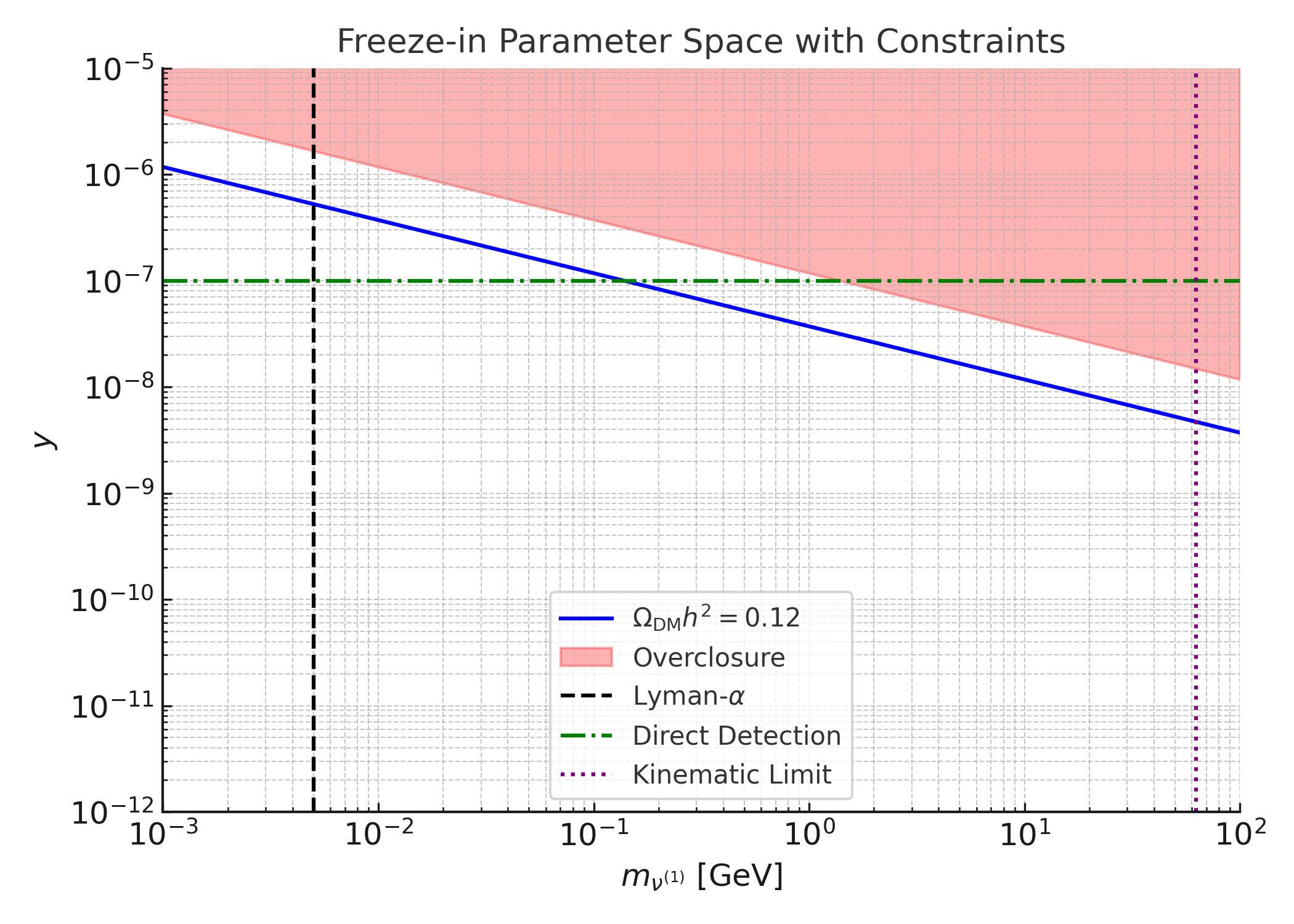}
    \caption{Contour showing the freeze-in relic density of the lightest sterile KK mode $\nu^{(1)}$ in the $(y, m_{\nu^{(1)}})$ plane. The blue curve yields the observed dark matter relic abundance $\Omega_{\rm DM} h^2 \simeq 0.12$. The red-shaded region overproduces dark matter and is excluded. While masses up to $\mathcal{O}(10~\text{GeV})$ are shown for generality, only the region $m_{\nu^{(1)}} \lesssim 1~\text{MeV}$ is consistent with all cosmological bounds (see Eq.~\ref{limit}).
        \textbf{Left:} The blue line shows the coupling \( y \) required to match the observed relic abundance. 
        The red shaded region indicates overclosure. 
        \textbf{Right:} Same as left, with additional constraints from Lyman-\(\alpha\) structure formation bounds (black dashed), 
        direct detection (green dash-dotted), and kinematic limits from Higgs decays (purple dotted).
    }
    \label{fig:freezein}
\end{figure*}

The left panel of Fig.~\ref{fig:freezein} displays the contour in the $(y, m_{\nu^{(1)}})$ plane for which the freeze-in production of $\nu_R^{(1)}$ yields the correct dark matter relic abundance. The region above this line (shaded red) leads to an overproduction of dark matter and is excluded by overclosure constraints. For small masses ($m_{\nu^{(1)}} \ll m_H$), the phase space for Higgs decay remains open, and the production is efficient even for extremely suppressed couplings $y \sim 10^{-10}$. As the mass increases toward the Higgs threshold, Boltzmann suppression becomes significant, and larger couplings are required to compensate. However, increasing $y$ also enhances the production rate, eventually pushing the relic density above the observed value and violating cosmological constraints.

The freeze-in mechanism is particularly well suited to light, feebly coupled states that remain out of equilibrium. For this reason, the region of interest is sharply localized around $m_{\nu^{(1)}} \sim \text{keV--MeV}$ and $y \sim 10^{-10}$, where the model remains predictive and consistent with cosmological observations. While Fig.~\ref{fig:freezein} spans a broad mass range up to $\mathcal{O}(10~\text{GeV})$ for completeness, it is important to emphasize that only the lower portion of this parameter space—specifically, $m_{\nu^{(1)}} \lesssim 1~\text{MeV}$—is compatible with all observational bounds. In particular, heavier sterile neutrinos are excluded by a combination of overproduction constraints, the Lyman-$\alpha$ forest data, and the kinematic limit for Higgs decay. The extended mass range in the figure is presented to illustrate the full behavior of the freeze-in contour across many orders of magnitude, but the viable and cosmologically consistent region remains sharply confined to the sub-MeV range.

In the right panel of Fig.~\ref{fig:freezein}, we overlay key observational constraints. Lyman-$\alpha$ forest data disfavors dark matter candidates with excessive free-streaming lengths, which places a lower bound on $m_{\nu^{(1)}}$. In parallel, bounds on $\Delta N_{\rm eff}$ limit the energy density injected by the decays of heavier KK modes into relativistic Majorons. Additional constraints come from the Higgs decay kinematics, which forbid production of sterile states heavier than $m_H/2$, and from experimental searches for exotic Higgs decays and neutrino interactions. Taken together, these bounds exclude much of the nominal parameter space and carve out a narrow allowed band that is sharply delineated in Fig.~\ref{fig:freezein}.

Finally, accounting for volume suppression in the effective couplings due to extra-dimensional propagation, the preferred region corresponds to $y \sim 10^{-10}$ and $g \sim 10^{-6}$. This degree of suppression is naturally achieved in large extra dimension models, and it enables a successful realization of WDM production and dark radiation injection within a unified framework. The freeze-in abundance, relic constraints, and radiative signatures of the KK tower together render the scenario highly predictive and falsifiable with future data from CMB-S4, Lyman-$\alpha$ surveys, and neutrino telescopes.

\section{Boltzmann System and \texorpdfstring{$\Delta N_{\rm eff}$}{Delta Neff}}

The decay of heavy sterile KK modes, $\nu_R^{(n)}$, into active neutrinos and relativistic Majorons, $\nu^{(n)} \to \nu + J$, injects additional radiation into the Universe after freeze-out. Unlike thermal relics, this radiation is nonthermal and arises after neutrino decoupling, leading to a shift in the effective number of relativistic degrees of freedom, $\Delta N_{\rm eff}$. To accurately track this contribution, we solve the coupled Boltzmann equations describing the evolution of the energy densities of the sterile neutrinos $\rho_{\nu^{(n)}}$ and the emergent Majoron population $\rho_J$:
\begin{align}
\frac{d\rho_J}{dt} + 4H\rho_J &= \sum_n \Gamma_n \rho_{\nu^{(n)}}, \\
\frac{d\rho_{\nu^{(n)}}}{dt} + 3H \rho_{\nu^{(n)}} &= -\Gamma_n \rho_{\nu^{(n)}},
\end{align}
where $\Gamma_n$ is the decay width of the $n$-th KK mode, and $H$ is the Hubble parameter. The initial conditions $\rho_{\nu^{(n)}}(T)$ depend on the sterile production history—either thermal or freeze-in.

The radiation injected by these decays contributes to the total radiation density at photon decoupling and is interpreted as an effective increase in the number of relativistic neutrino species. The associated shift in $\Delta N_{\rm eff}$ is given by
\begin{equation}
\Delta N_{\rm eff} = \frac{8}{7} \left(\frac{11}{4}\right)^{4/3} \left. \frac{\rho_J}{\rho_\gamma} \right|_{\rm CMB},
\end{equation}
where $\rho_\gamma$ is the photon energy density at recombination.

\begin{figure}[t]
    \centering
    \includegraphics[width=0.48\textwidth]{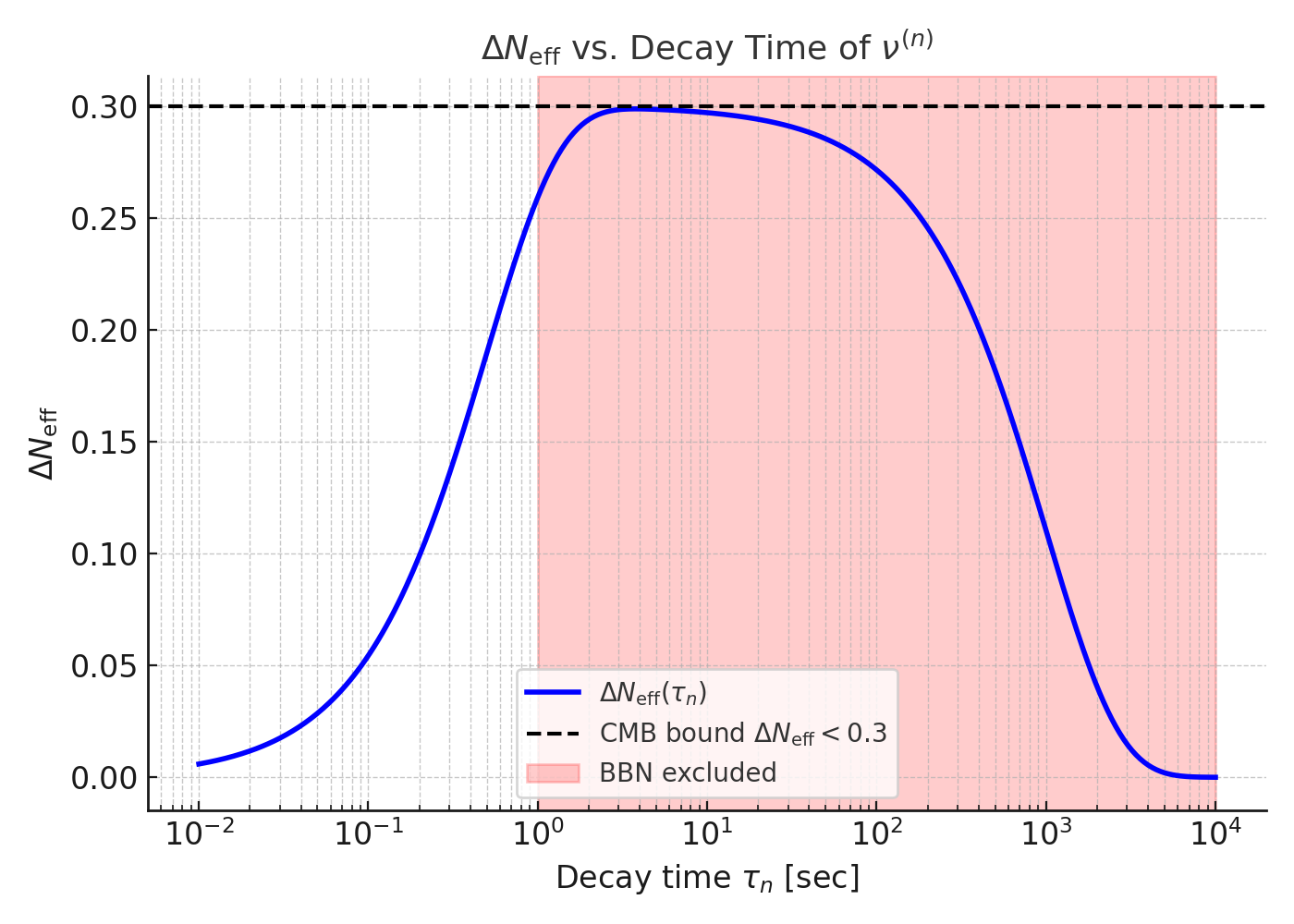}
    \caption{Predicted $\Delta N_{\rm eff}$ as a function of the decay time $\tau_n$ of sterile KK modes. Early decays are thermalized with the SM plasma, while late decays violate BBN constraints. The viable window lies around $\tau_n \sim 0.1$--$1$~s. These results apply across a broad KK mass range; however, consistency with dark matter and decay constraints favors $m_{\nu^{(1)}} \lesssim 1~\text{MeV}$.}
    \label{fig:deltaNeff}
\end{figure}

We solve the above system numerically using representative values of $g^{(n)}$, $m_n$, and initial comoving yields. As illustrated in Fig.~\ref{fig:deltaNeff}, the timing of KK mode decays is crucial. If decays occur sufficiently early ($\tau_n \ll 1$~s), the injected energy is redistributed within the SM plasma, saturating $\Delta N_{\rm eff}$ at the thermalization limit. Conversely, if decays occur too late ($\tau_n \gg 1$~s), they disturb Big Bang Nucleosynthesis (BBN), altering the predicted light element abundances. This defines a narrow temporal window, typically around $\tau_n \sim 0.1$--$1$~s, where sterile neutrino decays produce nonthermal radiation consistent with all cosmological bounds.

This mechanism enables a predictive cosmological signature directly tied to the model's freeze-in dynamics and decay structure, offering a means to probe extra-dimensional physics through precision measurements of $\Delta N_{\rm eff}$ in future CMB surveys.

\section{Neutrino Decay}

A further consequence of the Majoron coupling is the possibility of invisible decay of active neutrinos, $\nu_i \to \nu_j + J$, arising from the spontaneous breaking of the global lepton number symmetry. The decay width for this process is
\begin{equation}
\Gamma_{\nu_i} = \frac{|g_{ij}|^2}{16\pi} m_{\nu_i},
\end{equation}
leading to a rest-frame lifetime of
\begin{equation}
\tau_{\nu_i} \simeq 1.3 \times 10^{28} \left( \frac{10^{-6}}{g} \right)^2 \left( \frac{0.1~\text{eV}}{m_{\nu_i}} \right)\,\text{s}.
\end{equation}

For light neutrinos with $m_{\nu_i} \sim \mathcal{O}(\text{eV})$, strong bounds from solar, atmospheric, and astrophysical neutrino data impose $\tau_{\nu_i}/m_{\nu_i} \gtrsim 10^9~\text{s/eV}$~\cite{Beacom:2002vi, Bustamante:2020mep, Berryman:2022hkn}, implying $g \lesssim 10^{-6}$. These limits are especially relevant for long-baseline experiments and high-energy neutrino observatories such as IceCube.

In the heavier mass regime ($m_{\nu_i} \gtrsim \text{MeV}$), laboratory experiments and cosmological considerations impose complementary constraints. These include limits from meson decays, invisible $Z$ decays, and alterations to BBN and cosmic expansion history. Fast decays in this mass range can disrupt well-tested early-universe observables unless strongly suppressed.

\begin{figure}[t]
    \centering
    \includegraphics[width=0.48\textwidth]{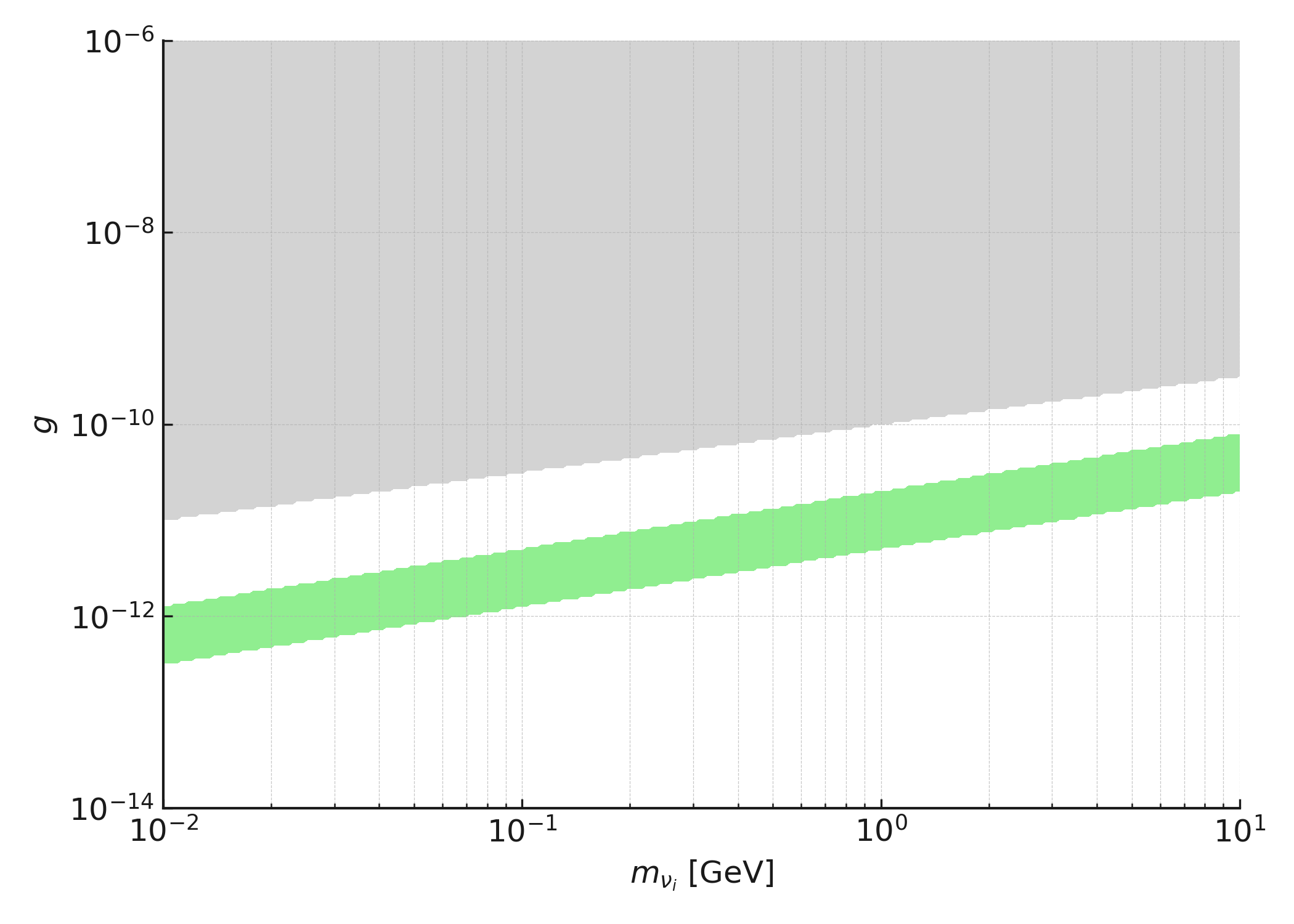}
    \caption{Constraints on the invisible neutrino decay coupling $g$ as a function of neutrino mass $m_{\nu_i}$. The gray region is excluded by neutrino decay bounds from astrophysical and laboratory data. The green band denotes the parameter space consistent with both dark radiation and freeze-in dark matter. Only the sub-MeV mass range remains viable across all constraints.}
    \label{fig:neutrinoDecay}
\end{figure}

Figure~\ref{fig:neutrinoDecay} summarizes these constraints in the $(g, m_{\nu_i})$ parameter space. The shaded gray region is excluded by lifetime bounds and collider or cosmological searches. The green band highlights the viable region where freeze-in production of $\nu_R^{(1)}$, successful $\Delta N_{\rm eff}$ generation, and allowed neutrino decay rates coexist. Notably, the sub-MeV regime simultaneously satisfies all three sectors—dark matter, dark radiation, and neutrino stability.

\section{Constraints and Predictions}

The model’s parameter space is shaped by a rich interplay between cosmological, astrophysical, and laboratory constraints. Limits on the effective number of relativistic species from Planck and BBN data constrain the radiation density to $\Delta N_{\rm eff} < 0.3$, placing an upper bound on the energy injection from KK mode decays. To satisfy this, sterile decays must occur before neutrino decoupling ($T \gtrsim 1~\text{MeV}$), translating to lifetimes $\tau_n \lesssim 1~\text{s}$.

Warm dark matter limits from Lyman-$\alpha$ observations impose a lower mass bound of $m_{\nu^{(1)}} \gtrsim 2~\text{keV}$ to prevent excessive free-streaming. Simultaneously, the freeze-in mechanism requires that $\nu_R^{(1)}$ never thermalizes, enforcing an upper limit on the Yukawa coupling: $y \lesssim 10^{-10}$, or equivalently $g \lesssim 10^{-6}$, once volume suppression is accounted for. These conditions carve out a sharply constrained region in the keV–MeV range.

Invisible decays of active neutrinos, $\nu_i \to \nu_j + J$, must comply with solar, atmospheric, and astrophysical data. For light ($\lesssim$~eV) states, lifetime bounds of $\tau_\nu / m_\nu \gtrsim 10^9$~s/eV again impose $g \lesssim 10^{-6}$. Heavier sterile neutrinos in the MeV–GeV regime face complementary bounds from beam dump experiments, meson decay studies, and constraints on nonstandard energy loss in supernovae.

Additionally, supernova cooling arguments exclude excessive Majoron emission for light $m_J$, especially in models with light active neutrinos and strong coupling. However, these bounds weaken for heavier sterile states or suppressed couplings.

Taking all constraints into account, the allowed parameter space is both predictive and testable:
\begin{align} \label{limit}
    R^{-1} &\sim 10^3~\mathrm{TeV}, \quad g \lesssim 10^{-6}, \nonumber \\
    m_{\nu^{(1)}} &\sim 2~\mathrm{keV} - \mathcal{O}(\mathrm{MeV}), \quad
    m_{\nu_i} &\gtrsim \mathrm{eV}~\text{to}~\mathrm{GeV}.
\end{align}
Although Figs.~\ref{fig:freezein}--\ref{fig:neutrinoDecay} display an extended mass range for generality, only the sub-MeV window—particularly $m_{\nu^{(1)}} \sim 2~\mathrm{keV}$ to $1~\mathrm{MeV}$—satisfies all observational constraints and unifies the production of dark matter, dark radiation, and neutrino decay through the same underlying Majoron portal.

This sharply defined regime offers promising experimental targets for next-generation probes such as CMB-S4, Simons Observatory, DUNE, IceCube-Gen2, and high-resolution Lyman-$\alpha$ forest data. Any observed deviation in $\Delta N_{\rm eff}$ or detection of neutrino decay would provide smoking-gun signatures of this unified framework.

\section{Conclusion}

We have proposed a unified and minimal framework in which right-handed neutrinos propagating in large extra dimensions interact with a pseudo-Nambu--Goldstone boson (Majoron) arising from spontaneous lepton number violation. This construction simultaneously accounts for three key phenomena beyond the Standard Model: the nature of dark matter, the presence of excess dark radiation, and the possibility of invisible neutrino decay.

In this scenario, the lightest sterile Kaluza--Klein (KK) mode is produced via the freeze-in mechanism through suppressed Higgs decays and serves as a warm dark matter candidate. Heavier KK modes decay to relativistic Majorons and active neutrinos, producing a nonthermal contribution to $\Delta N_{\rm eff}$, while the Majoron also mediates invisible decays of active neutrinos. These processes are tightly connected through the extra-dimensional volume suppression and a common interaction structure.

Although Figs.~\ref{fig:freezein}--\ref{fig:neutrinoDecay} present results over a broad parameter range, including sterile neutrino masses up to $\mathcal{O}(10~\text{GeV})$, we emphasize that only the sub-MeV regime—specifically $m_{\nu^{(1)}} \sim 2~\mathrm{keV}$–$1~\mathrm{MeV}$—is compatible with all cosmological and laboratory constraints. The extended mass range is included to illustrate the full shape of the freeze-in contour and the behavior of the decay-induced radiation injection, but viable solutions lie exclusively within the light sterile region. This emphasizes the predictive nature of the framework.

The model points to a narrow and testable region of parameter space, with characteristic values $g \sim 10^{-6}$ and compactification scale $R^{-1} \sim 10^3~\mathrm{TeV}$. It offers distinctive cosmological signatures—including sub-MeV dark matter, correlated contributions to $\Delta N_{\rm eff}$, and suppressed neutrino lifetimes—that can be probed in the near future. Experiments such as CMB-S4, Simons Observatory, DUNE, IceCube-Gen2, and precision Lyman-$\alpha$ surveys will be sensitive to these effects, allowing this framework to be tested or ruled out across multiple frontiers.

\section{Acknowledgment}
The work of AC was supported by the Japan Society for the Promotion of Science (JSPS) as a part of the JSPS Postdoctoral Program (Standard), grant number JP23KF0289.

\end{document}